\documentclass[bibyear]{aa}  

\usepackage{graphicx}
\usepackage{txfonts}
\begin{document}

   \title{An ancient F-type subdwarf from the halo crossing 
          the Galactic plane\thanks{Based on observations at the 
La Silla-Paranal Observatory of the European Southern Observatory 
for programmes 092.D-0040(A) and 093.D-0127(A).}}


\titlerunning{An ancient F-type subdwarf}

   \author{R.-D. Scholz\inst{1}
          \and
          U. Heber\inst{2}
          \and
          C. Heuser\inst{2}
          \and
          E. Ziegerer\inst{2} 
          \and
          S. Geier\inst{3}
          \and
          F. Niederhofer\inst{3}
          }

   \institute{Leibniz Institute for Astrophysics Potsdam (AIP),
              An der Sternwarte 16, 14482 Potsdam, Germany
              \email{rdscholz@aip.de}
              \and
              Dr. Remeis Observatory \& ECAP, Astronomical Institute,
              Friedrich-Alexander University Erlangen-N\"urnberg,
              Sternwartstr. 7, 96049 Bamberg, Germany
              \and
              European Southern Observatory, 
              Karl-Schwarzschild-Str. 2, 85748 Garching, Germany
             }

   \date{Received 5 December 2014; accepted 19 December 2014}

 
  \abstract
   {}
   {We selected the bluest object, WISE~J0725$-$2351, 
    from Luhman's new high proper
    motion (HPM) survey based on observations with the
    Wide-field Infrared Survey Explorer (WISE) for spectroscopic follow-up
    observations. 
    Our aim was to unravel the nature of this relatively
    bright ($V$$\sim$12, $J$$\sim$11) 
    HPM star ($\mu$$=$267\,mas/yr).}
   {We obtained low- and medium-resolution spectra with the 
    European Southern Observatory (ESO) 
    New Technology Telescope (NTT)/EFOSC2 and 
    Very Large Telescope (VLT)/XSHOOTER instruments, 
    investigated the radial velocity 
    and performed a quantitative spectral analysis that allowed us to determine
    physical parameters. The fit of the spectral energy distribution 
    based on the available photometry to low-metallicity model spectra
    and the similarity of our target to a metal-poor benchmark star (HD~84937)
    allowed us to estimate the distance and space velocity.}
   {As in the case of HD~84937, we classified WISE~J0725$-$2351 as sdF5: 
    or a metal-poor turnoff star with
    $[Fe/H]$$=$$-$2.0$\pm$0.2, 
    $T_{eff}$$=$6250$\pm$100\,K,
    $\log{g}$$=$4.0$\pm$0.2, 
    and a possible age of about 12\,Gyr. At an estimated
    distance of more than 400\,pc, its proper motion translates to a
    tangential velocity of more than 500\,km/s. Together with
    its constant (on timescales of hours, days, and months) and 
    large radial velocity (about $+$240\,km/s), the resulting
    Galactic restframe velocity
    is about 460\,km/s, 
    implying a bound retrograde orbit for this 
    extreme
    halo object that currently crosses the Galactic plane
    at high speed.}
   {}


   \maketitle


\section{Introduction}
\label{Sintro}

Most of the roughly 50000 known high proper motion (HPM) stars 
(with $\mu$$\gtrsim$0.2\,arcsec/yr)
were detected in optical 
surveys based on photographic Schmidt plates (e.g. by
Luyten~\cite{luyten79a,luyten79b},
Scholz et al.~\cite{scholz00}, 
Pokorny et al.~\cite{pokorny04},
Hambly et al.~\cite{hambly04},
L{\'e}pine \& Shara~\cite{lepine05},
L{\'e}pine~\cite{lepine05a,lepine08}).
The majority of these HPM stars are not high velocity, but just nearby 
(2$\lesssim$$d$$\lesssim$50\,pc) stars that belong 
to the thin disc and the same spiral arm as the sun. However, among
the HPM stars there are also few thick disc and even fewer Galactic 
halo stars (subdwarfs) that are on 
average several times further away (10$\lesssim$$d$$\lesssim$250\,pc), but
cross the solar neighbourhood at high speeds (with tangential velocities
of up to several 100\,km/s). Therefore, the proper motion alone can only
provide a crude distance estimate. Nevertheless, it is often used as
a starting point in the search for unknown nearby stars.

The incompleteness of the census of nearby stars, dominated by M dwarfs, 
increases with distance, as demonstrated for the northern and
southern hemispheres, respectively
by L{\'e}pine \& Gaidos~(\cite{lepine13}) 
and Winters et al.~(\cite{winters14}). For the immediate solar neighbourhood,
the Research Consortium on Nearby Stars
(RECONS)\footnote{http://www.chara.gsu.edu/RECONS/} gives regular
updates for a high-quality 10\,pc sample. All systems in that sample
have trigonometric parallaxes with errors of less than 10\,mas. From 2000 to
2012, the numbers of M dwarfs in this sample increased from 198 to 248
(by 25\%), whereas the number of white dwarfs (WDs) rose from 18 to 20 
(by about 10\%). This progress was achieved thanks to RECONS and other
parallax programmes that concentrated on previously detected new HPM objects.
Adric Riedel and the RECONS team also provided the numbers of 
known systems and different types of objects in the 15.625 times larger 
volume of the 25\,pc sample in a video of 12 June 2014. The numbers of
M dwarfs (1093) and WDs (137) anounced in this video are two-three times 
smaller than the expected numbers ($\sim$3900 and $\sim$310, respectively), 
if one assumes that the number densities do not change from 10 to 25\,pc.
This demonstrates the potential for the identification of hitherto unknown
WD and M dwarf neighbours of the sun and the need for ongoing 
and new parallax programmes. Concerning nearby WDs, 
there is an interesting lack of moderately HPM objects: 
100\% of the known WDs within 10\,pc have total 
proper motions $>$500\,mas/yr and 84\% have $>$1000\,mas/yr, whereas for other 
stars these fractions are only 81\% and 49\%, respectively (data from SIMBAD).
We suspect that some nearby WDs with relatively small proper motions have not 
yet been identified. 

We note that only four subdwarfs 
(LHS~29, LHS~189, LHS~272, and LHS~406)
are included in the above-mentioned
video on the 25\,pc sample, and all these are M-type subdwarfs.
The late-M subdwarf SSSPM~J1444$-$2019 (Scholz et al.~\cite{scholz04})
with a parallax of 61.67$\pm$2.12\,mas
(Schilbach et al.~\cite{schilbach09}) should be added here.
Jao et al.~(\cite{jao08}) reported in their Table~2 on some additional 
subdwarfs possibly falling in the 25\,pc sample 
(with $K_s$$-$$M_{K_s}$$\lesssim$2; see also 
Table 4 of Winters et al.~\cite{winters14}). They
also mentioned (in their Sect.7.2.2)
the nearest subdwarf binary $\mu$ Cas~AB
(listed in SIMBAD with a spectral type of ''K1V\_Fe-2'')
with an accurate parallax of 132.38$\pm$0.82\,mas 
(van Leeuwen~\cite{vanleeuwen07}). 
Another K-type subdwarf binary possibly within 25\,pc is LHS~72/73 
(Reyl{\'e} et al.~\cite{reyle06}, 
Jao et al.~\cite{jao08},
Rajpurohit et al.~\cite{rajpurohit14}). As in the case of 
main-sequence dwarfs,
K- and especially M-type subdwarfs are much more frequent in the solar 
neighbourhood than earlier-type subdwarfs. The nearest known F-
and G-type subdwarfs with trigonometric parallaxes lie according to SIMBAD
slightly beyond 50\,pc. Concerning A-type subdwarfs, there is no clear evidence
for their existence. Besides some historical papers on this topic
(Chamberlain \& Aller~\cite{chamberlain51}, Greenstein~\cite{greenstein54}),
we found only two A-type subdwarfs with trigonometric parallaxes in SIMBAD
(corresponding to distances $>$180\,pc).
One of those (HD~224927) is a close binary with an A-type primary, 
whereas the other (HD~161817) is classified as a horizontal branch star. 
Their sdA8 and sdA2 types, respectively listed in SIMBAD, are both outdated
as can be seen in the General Catalogue of Stellar Spectral Classifications
(Skiff~\cite{skiff14}).

The cool subdwarf sequence currently reaches from moderately 
cool F- to ultracool 
late-M-, L- and T-types and represents a completely different class of objects 
than hot O- and B-type 
subdwarfs. Their domains in a Hertzsprung–Russell diagram are clearly 
separated, as e.g. shown by Gontcharov et al.~(\cite{gontcharov11}), who 
called them unevolved and evolved subdwarfs, respectively. 
Jao et al.~(\cite{jao08}) suggested using the ''sd'' prefix only for
the hot evolved subdwarfs (sdO, sdB) and identifying all the cool unevolved
subdwarfs by their luminosity class ''VI''. 
Drilling et al.~(\cite{drilling13}) presented a three-dimensional
spectral classification (spectral, luminosity, and helium class) for the 
hot sdO and sdB subdwarfs.
Over the last few decades, new
classification systems were developed and refined for K- and M-type
subdwarfs, with decreasing metallicity from normal subdwarfs (sd), to
extreme (esd), and ultra (usd) subdwarfs 
(Gizis~\cite{gizis97}, L{\'e}pine, Rich \& Shara~\cite{lepine07}).
Many new ultracool subdwarfs have been discovered in recent years
(see e.g. review by 
Burgasser et al.~\cite{burgasser09},
Cushing et al.~\cite{cushing09},
Lodieu et al.~\cite{lodieu12},
Zhang et al.~\cite{zhang13},
Wright et al.~\cite{wright14},
Kirkpatrick et al.~\cite{kirkpatrick14},
Luhman \& Sheppard~\cite{luhman14}),
whereas the warm end of the cool subdwarf sequence was not
in the main focus of research (in terms of new discoveries and spectroscopic
classification schemes).

Here, we report the identification and classification of a new F-type
subdwarf crossing the Galactic plane at high speed.
We selected this target as a rather blue and bright object in a
new HPM survey and first suspected it to be a very nearby WD.
(Sect.~\ref{Starget}). In Sect.~\ref{Spm} and \ref{Sphoto}
we present our improved proper motion solution and the
collected photometric data, respectively.
Follow-up spectroscopic observations 
and radial velocity measurements are described in Sect.~\ref{Sspecrv},
whereas Sect.~\ref{Sqspec} deals with the 
spectral analysis.
The distance and
kinematics of the new halo object are estimated in Sect.~\ref{Sdistv}.
In Sect.~\ref{Sdiscu} we give our conclusions and a brief discussion of 
our results.

   \begin{figure}
   \centering
   \includegraphics[width=8.8cm]{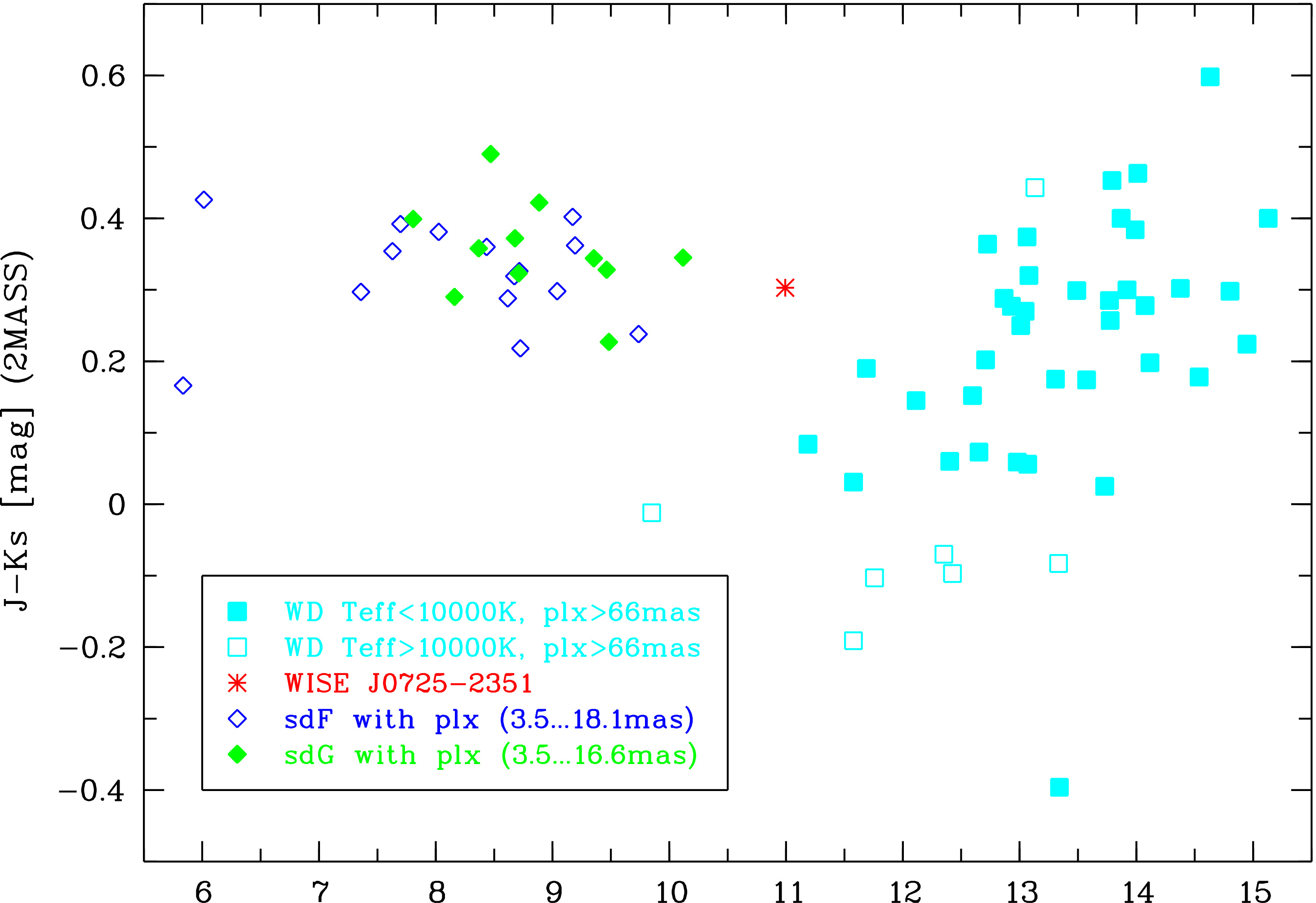}
   \includegraphics[width=8.8cm]{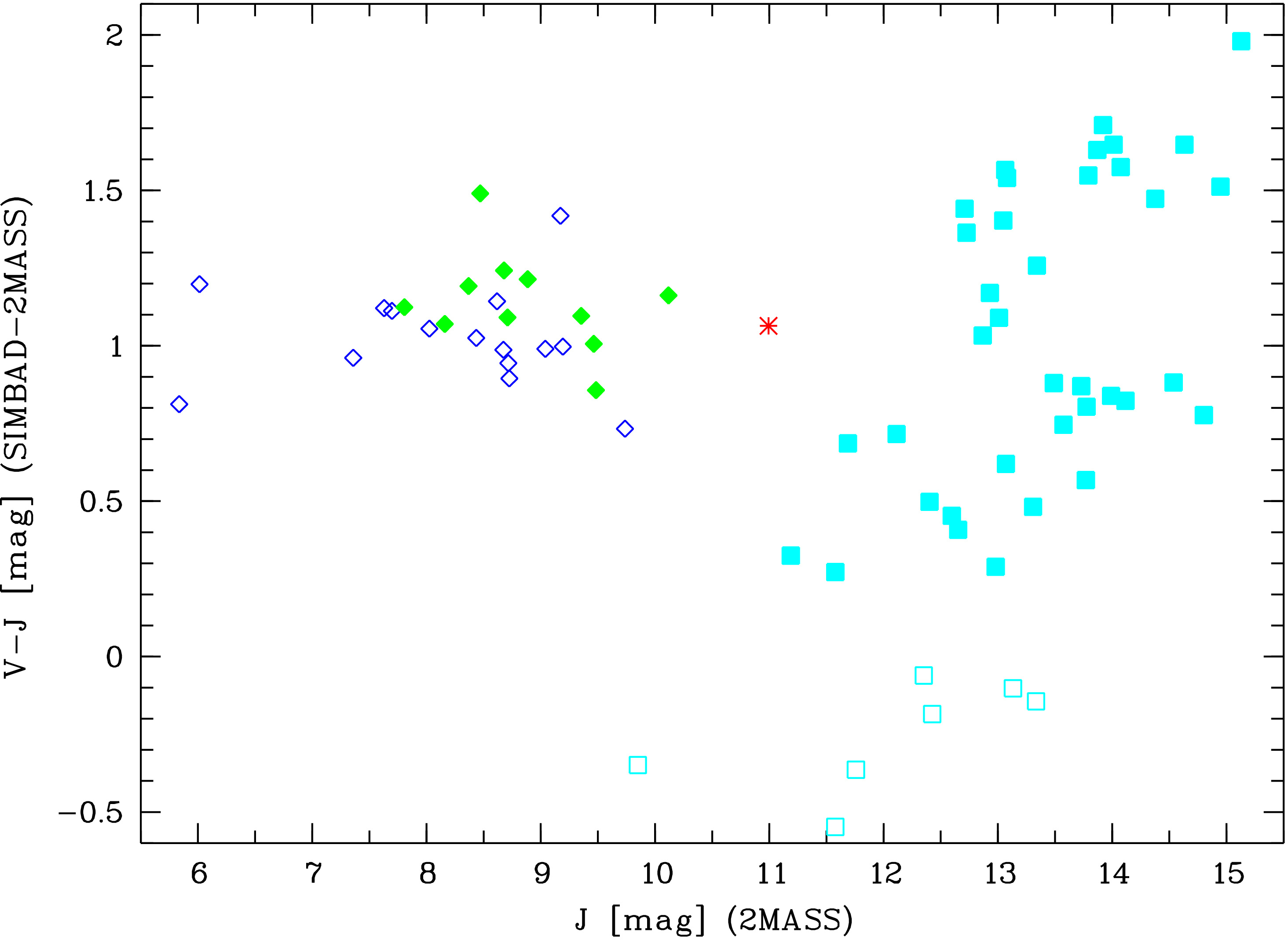}
      \caption{\textbf{Top:} Near-infrared colour-magnitude diagram showing  
               WISE~J0725$-$2351 in comparison to nearby WDs and
               sdF/sdG subdwarfs with trigonometric parallaxes and
               available 2MASS photometry (close binaries excluded).
               \textbf{Bottom:} 
               Optical-to-near-infrared colour-magnitude diagram
               showing the same objects.
              }
         \label{Fjksv}
   \end{figure}


\section{Target selection}
\label{Starget}

Multiple epochs from the
Wide-field Infrared Survey Explorer (WISE; Wright et al.~\cite{wright10}), 
supplemented by about ten years older data from the Two Micron All Sky Survey
(2MASS; Skrutskie et al.~\cite{skrutskie06}),
served as the basis for two new infrared HPM surveys (Luhman~\cite{luhman14a};
Kirkpatrick et al.~\cite{kirkpatrick14}) independent of photographic
Schmidt plates. Both surveys aimed at the discovery of
very nearby cool brown dwarfs 
(Luhman~\cite{luhman13,luhman14a,luhman14b}) and new L-type subdwarfs
(Wright et al.~\cite{wright14},
Kirkpatrick et al.~\cite{kirkpatrick14},
Luhman \& Sheppard~\cite{luhman14}).
Among the new HPM objects of the Luhman~(\cite{luhman14a}) sample, there
are also some blue objects, obviously overlooked in previous HPM surveys
based on Schmidt plates. Among theses sources, we selected the object with
the AllWISE (Kirkpatrick et al.~\cite{kirkpatrick14}) designation
WISE~J072543.88-235119.7 (hereafter WISE~J0725$-$2351), also
known as 2MASS J07254392$-$2351168, which had the
smallest colour indices $J$$-$$w2$$=$0.35 and $J$$-$$K_s$$=$0.30,
according to the 2MASS and WISE all-sky catalogue.

With these available colours and the bright magnitude ($J$$\sim$11.0)
we considered WISE~J0725$-$2351 as a very nearby WD candidate and initiated
spectroscopic follow-up observations 
(Sect.~\ref{Sspecrv}). 
However, we also mentioned that sdF and sdG subdwarfs have similar colours. 
In Fig.~\ref{Fjksv} (top) we compare the $J$$-$$K_s$ colour and $J$ magnitude
of our target with those of the known WDs within 15\,pc (close
binaries were excluded) and of all known sdF and sdG subdwarfs with
available parallaxes as provided by SIMBAD. 
We are aware that these subdwarf samples may be not complete and may be
contaminated with stars of different spectral and luminosity classes because 
of ambiguous classification or missing updates in SIMBAD.
However, for both sdF and sdG in SIMBAD,
their parallaxes range between about 3.5\,mas and 19\,mas, corresponding
to distances between about 50\,pc and 300\,pc, where the relative errors
of the smaller parallaxes are very large. All but one of the WDs are fainter
than WISE~J0725$-$2351, whereas all the sdF and sdG with measured parallaxes
are brighter. The $J$$-$$K_s$ colour of our target is consistent with that
of the sdF, sdG, and the cool WDs. As we later found additional photometry
for WISE~J0725$-$2351 (Sect.~\ref{Sphoto}), we include a $V$$-$$J$,$J$ 
diagram at the bottom of Fig.~\ref{Fjksv}, where our target is again located
between the regions occupied by sdF/sdG and cool WDs.

   \begin{figure}
   \centering
   \includegraphics[width=8.8cm]{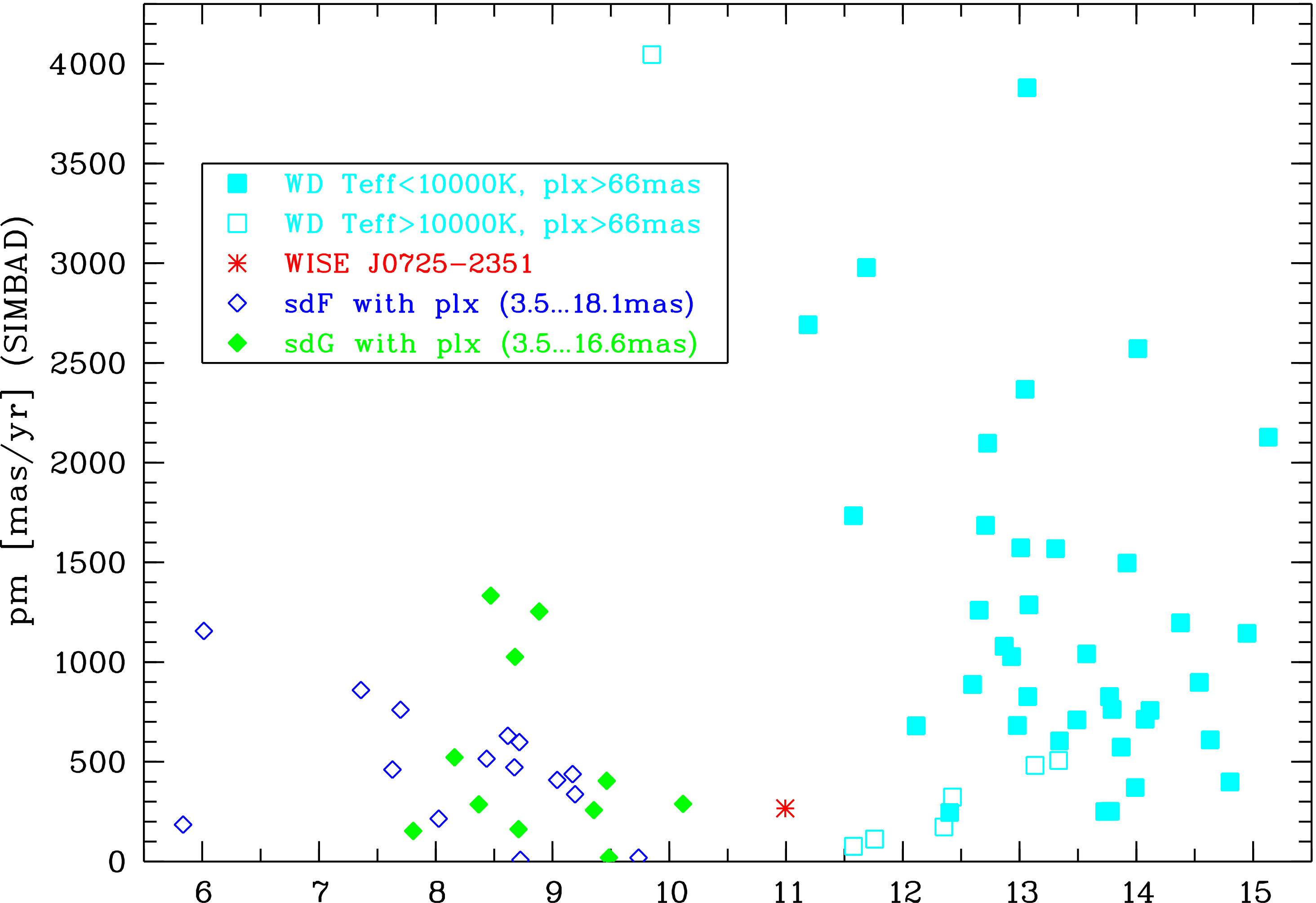}
   \includegraphics[width=8.8cm]{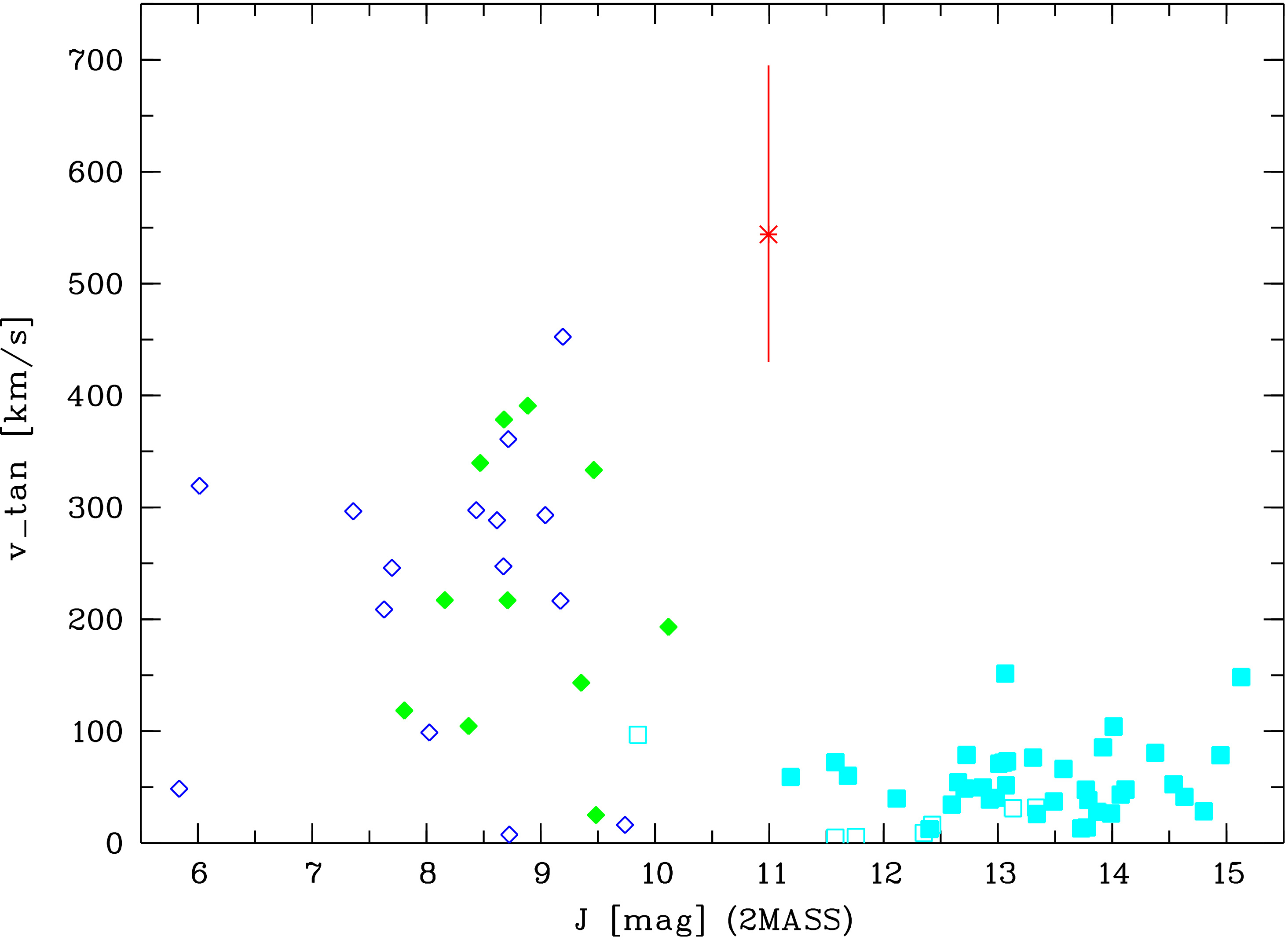}
      \caption{\textbf{Top:} Proper motion as a function of $J$ magnitude
               for WISE~J0725$-$2351, nearby WDs, and sdF/sdG subdwarfs.
               \textbf{Bottom:} Tangential velocities based on the photometric 
               distance estimate for WISE~J0725$-$2351 (Sect.~\ref{Sdistv})
               and the trigonometric parallaxes of all other objects
               (same objects and symbols as in Fig.~\ref{Fjksv}).
               For clarity, the error bars are shown for 
               WISE~J0725$-$2351 only.
               The error bars 
               of sdF/sdG subdwarfs may be even larger in cases of
               uncertain parallaxes, whereas for WDs they are typically
               comparable to the symbol size.
              }
         \label{Fjpmv}
   \end{figure}

   \begin{figure*}
   \sidecaption
   \includegraphics[width=12.9cm]{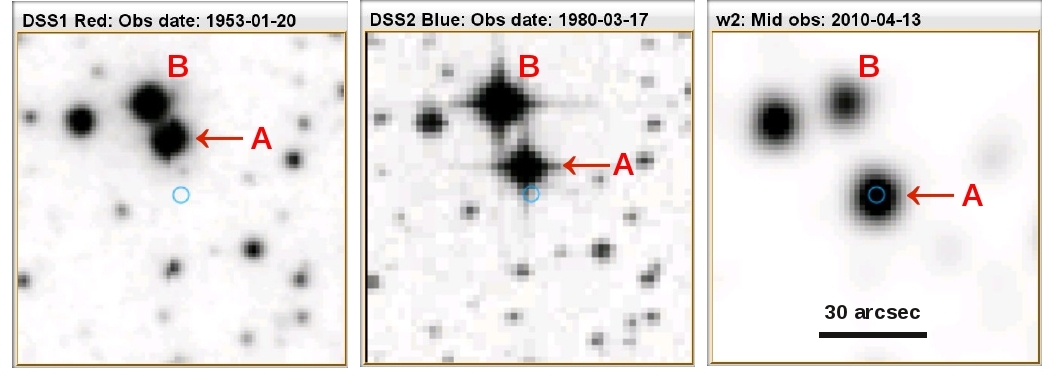}
      \caption{Finder charts of 90$\times$90\,arcmin$^2$ 
               (north is up, east to the left) from red and blue
               (first and second epoch, respectively) plates of the 
               Digitized Sky Surveys (DSS) and from the
               WISE $w2$-band centred on the position of
               our HPM star WISE~J0725$-$2351 (marked as object A) 
               in the WISE all-sky catalogue (blue circle).
               At earlier epochs, our target moved very close
               to a blue background star (object B; see text).
              }
         \label{Ffc3}
   \end{figure*}


\section{Improved proper motion}
\label{Spm}

The relatively bright star WISE~J0725$-$2351 was listed as a new
HPM object in Luhman~(\cite{luhman14a}). It was not detected in any 
previous HPM survey. However, the 
PPM-Extended (PPMX; R\"oser et al.~\cite{roeser08}) catalogue included 
this object with a somewhat different proper motion (Table~\ref{Tpm}),
the United States Naval Observatory (USNO) 
B1.0 (Monet et al.~\cite{monet03}) associates a zero
proper motion to this object,
the Southern Proper Motion (SPM4; Girard et al.~\cite{girard11}) catalogue
does not include this object,
whereas the Fourth US Naval Observatory CCD Astrograph Catalog (UCAC4; 
Zacharias et al.~\cite{zacharias13}) gives only an accurate position
but no proper motion. The reason for these discrepancies is probably
source confusion with a similarly bright blue background star
(hereafter object B), with
which our HPM star (object A) was almost overlapping at earlier epochs (see
Fig.~\ref{Ffc3}). Whereas both stars were resolved in the Tycho input
catalogue (Egret et al.~\cite{egret92}), only the background object B entered
the Tycho (ESA~\cite{esa97}) and Tycho-2 (H{\o}g et al.~\cite{hog00})
catalogues (as TYC1~TYC2~TYC3 = 6538~2171~1, with non-significant
proper motion components in Tycho-2 of less than 5\,mas/yr). 
Both stars were included
as a visual double star in the Washington double stars catalogue
(Mason et al.~\cite{mason01}) with different separations (from 3.7 to 
4.1\,arcsec) and position angles (from 256$^{\circ}$ to 247$^{\circ}$) 
measured between 1910 and 1922. This apparent double star
was also catalogued as CD~$-$23~5447 in the Cordoba 
Durchmusterung 
(Thome~\cite{thome1892})\footnote{http://cdsarc.u-strasbg.fr/viz-bin/Cat?I/114}.
Note that SIMBAD provides only one entry, CD~$-$23~5447, within a few arcmin of 
the Tycho-2 star, but places it, probably  because of the uncertain old input 
coordinates in the Cordoba Durchmusterung, at 07~25~44 $-$23~51.3 (ICRS coord),
very close (within a few arcsec) to the current position of our HPM star.

For our improved proper motion solution (Table~\ref{Tpm}), we combined
14 available multi-epoch positions: 
the two mean WISE positions from 2010 (from AllWISE) and the
2MASS position from 1999 used by Luhman~(\cite{luhman14a}),
the position from 1916 given in the Astrographic Catalogue 
(AC2000; Urban et al.~\cite{urban98}),
our own visual measurement of the Digitized Sky Survey red plate from 1953
and the blue plate from 1980,
five positions measured in
the SuperCOSMOS Sky Survey (SSS; Hambly et al.~\cite{hambly01a,hambly01c})
and the SuperCOSMOS H$_{\alpha}$ survey (Parker et al.~\cite{parker05})
with epochs from 1980 to 2002,
the 1999 position from UCAC4,
the 2006 position from
the last issue of the Carlsberg Meridian Catalogue (CMC15; 
Mui{\~n}os \& Evans~\cite{muinos14}),
and the 2012 position measured by the Galaxy Evolution Explorer 
(GALEX\footnote{http://galex.stsci.edu/GR6/?page=mastform}; 
Morrissey et al.~\cite{morrissey07}). The resulting proper motion of 
($\mu_{\alpha}\cos{\delta}$, $\mu_{\delta}$) = ($-$51.2$\pm$1.7, $-$261.8$\pm$0.8) is about two times more precise than the typical Tycho-2 proper motions
of similarly bright stars in the field around WISE~J0725$-$2351.

%
\begin{table}
\caption{Proper motion of WISE~J0725$-$2351} 
\label{Tpm}      
\centering                          
\begin{tabular}{lcc}        
\hline\hline                 
Source   & $\mu_{\alpha}\cos{\delta}$ & $\mu_{\delta}$   \\ 
         & [mas/yr]                   & [mas/yr] \\
\hline                        
Luhman~(\cite{luhman14a}) & $-$50$\pm$10 & $-$260$\pm$10 \\
AllWISE                   & $-$14$\pm$36 & $-$276$\pm$36 \\
USNO B1.0                 & 0            & 0             \\
PPMX                      & $-$74.1$\pm$3.1 & $-$265.3$\pm$2.4 \\
UCAC4                     & -               & -                \\
this work                 & $-$51.2$\pm$1.7 & $-$261.8$\pm$0.8 \\
\hline                                   
\end{tabular}
\end{table}

As seen in Fig.~\ref{Fjpmv} (top), this proper motion represents a relatively
small value for a nearby WD. If our target would have turned out to be a
WD, it would have been 
a cool WD (see Fig.~\ref{Fjksv}) within a few parsecs from
the sun (according to its relatively bright magnitude). Its proper motion 
would translate to a very small tangential velocity (few km/s). Though
this would be consistent with the majority of nearby WDs having tangential
velocities between 0 and 100\,km/s (Fig.~\ref{Fjpmv}, bottom).
However, the opposite classification of WISE~J0725$-$2351 as an 
F-type subdwarf (see Sect.~\ref{Sqspec}) and its relatively faint magnitude
with respect to the known objects of this class leads to a large distance
(Sect.~\ref{Sdistv}) and a very high tangential velocity (Fig.~\ref{Fjpmv}, 
bottom).


\section{Photometry}
\label{Sphoto}

We collected the photometric data
for WISE~J0725$-$2351 and of the blue background star
(object B in Fig.~\ref{Ffc3})
from VizieR\footnote{vizier.u-strasbg.fr/viz-bin/VizieR} at the 
Centre de Donn{\'e}es astronomiques de Strasbourg (CDS),
the American Association of Variable Star Observers (AAVSO) 
Photometric all-sky survey 
(APASS)\footnote{http://www.aavso.org/download-apass-data} 
Data Release 7, and GALEX, as presented in Table~\ref{Tphot}.
Note that the $r'$ magnitudes from APASS and CMC15 (and the $f.mag$
from UCAC4) are in very good agreement, respectively for both objects.
We have not included the photographic SSS photometry 
(Hambly et al.~\cite{hambly01a,hambly01b}) in Table~\ref{Tphot},
as these relatively bright stars are affected by saturation and image crowding
on the Schmidt plates.
 
%
\begin{table}
\caption{Photometry of WISE~J0725$-$2351 and background object B} 
\label{Tphot}      
\centering                          
\begin{tabular}{lcc}        
\hline\hline                 
Source/  & WISE~J0725$-$2351 & object B \\ 
band         & [mag]             & [mag]  \\
\hline                        
GALEX  $NUV$  &  15.05$\pm$0.01  &  13.81$\pm$0.01 \\
  \\
APASS $B$     & 12.482$\pm$0.035 & 11.521$\pm$0.035 \\
Tycho $BT$    & not available    & 11.239$\pm$0.088 \\
APASS $g'$    & 12.240$\pm$0.029 & 11.436$\pm$0.018 \\
APASS $V$     & 12.057$\pm$0.024 & 11.469$\pm$0.023 \\
Tycho $VT$    & not available    & 11.193$\pm$0.136 \\
UCAC4 $f.mag$ & 11.984           & 11.545                 \\
APASS $r'$    & 11.960$\pm$0.033 & 11.592$\pm$0.024 \\
CMC15 $r'$    & 11.919           & 11.549                 \\
APASS $i'$    & 11.819$\pm$0.018 & 11.716$\pm$0.005 \\
2MASS $J$     & 10.993$\pm$0.024 & 11.251$\pm$0.026       \\
2MASS $H$     & 10.720$\pm$0.022 & 11.228$\pm$0.022       \\
2MASS $K_s$   & 10.690$\pm$0.021 & 11.235$\pm$0.023       \\
AllWISE $w1$  & 10.616$\pm$0.023 & 11.184$\pm$0.024       \\
AllWISE $w2$  & 10.644$\pm$0.020 & 11.225$\pm$0.020       \\
AllWISE $w3$  & 10.612$\pm$0.081 & 11.262$\pm$0.148       \\
\hline                                   
\end{tabular}
\end{table}

With respect to our target WISE~J0725$-$2351, the background object B appears
blue, with all magnitudes from APASS $B$ to AllWISE $w3$ being in the range
of about 11.2 to 11.7, whereas WISE~J0725$-$2351 is nearly two magnitudes
brighter in the mid-infrared WISE bands compared to the APASS $B$-band.
According to its colour and small proper motion (Sect.\ref{Spm}), object B 
could be an A star at a distance of $\sim$700-1600\,pc (as derived from the 
comparison with the A7V star Altair and the A0V star Vega) with a tangential 
velocity of less than 16-38\,km/s, which is typical of a thin disc star.
Alternatively, object B could also be a B star (e.g. like the B7V 
star Regulus) but slightly reddened ($A_V$$\sim$0.2). In that case it would lie
at a distance of about 2500\,pc with a tangential velocity of less than 
60\,km/s, which would still be realistic.

As shown in Fig.~\ref{Fjksv} for the $V$$-$$J$ and $J$$-$$K_s$ colours, 
WISE~J0725$-$2351 appears similar to both sdF/sdG subdwarfs and cool WDs.
Its $g'$$-$$r'$$=$0.28 and $r'$$-$$i'$$=$0.14 are consistent with the
selection criteria for distant and much fainter ($r'$$>$15) halo F-type stars 
used in Allende Prieto et al.~(\cite{allende14}); and
with $B$$-$$V$$=$0.43, it is also similar to the nearby F5V star Procyon.
However, with the absolute $V$ magnitude of Procyon, WISE~J0725$-$2351
would be about 780\,pc away and consequently move at a tangential velocity of
about 990\,km/s, which is unlikely. An F-type subdwarf rather than a 
main-sequence star classification (Sect.~\ref{Sqspec}) leads to a smaller
distance and more realistic  velocity (Sect.~\ref{Sdistv}). 

   \begin{figure}
   \centering
   \includegraphics[width=5.6cm]{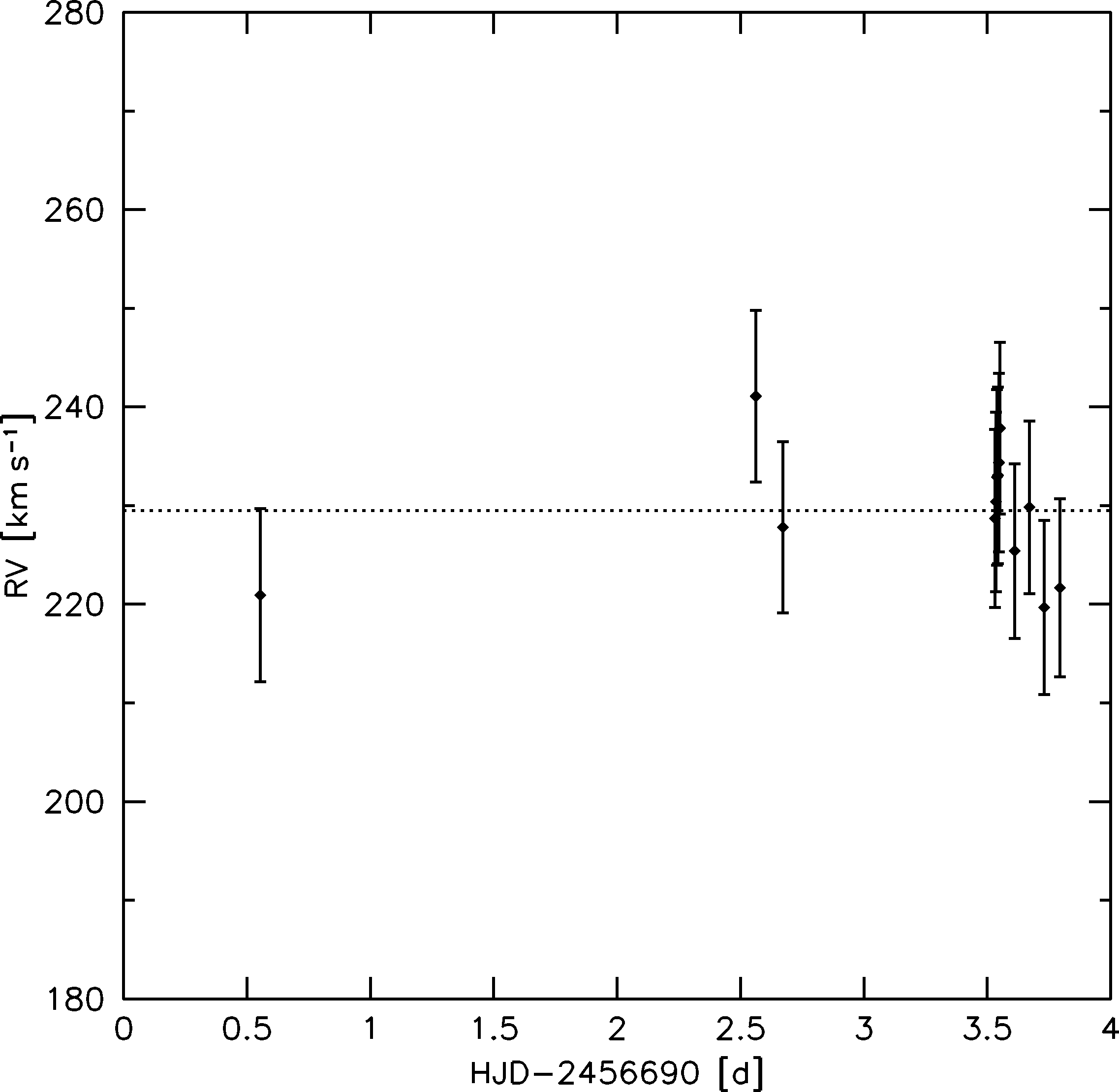}
      \caption{Radial velocity measurements of WISE~J0725$-$2351 from
               EFOSC2 spectra observed during three nights in
                February 2014. The dotted line shows
               the mean of these values.
               }
         \label{Frv}
   \end{figure}
 

\section{Spectroscopic observations, spectral classification, and radial velocity curve}
\label{Sspecrv}

To classify the star we obtained a low-resolution spectrum 
($R$$\simeq$700, $\lambda$$=$3300-5200\AA{}) with
the EFOSC2 spectrograph mounted at the 
European Southern Observatory (ESO) New Technology Telescope (NTT) 
in February 2014.
Reduction was done with standard MIDAS procedures.
The spectrum showed the Balmer series and very prominent Ca H\&K lines, 
which excluded a white dwarf.
To improve the classification and for a subsequent quantitative 
spectral analysis, we obtained another set of six single
spectra ($R$$=$5000-7500, $\lambda$$=$3000-25\,000\AA{}) 
with XSHOOTER mounted at the ESO Very Large Telescope (VLT) on May 8, 2014.
While those spectra have been taken with narrow slits 0.8-0.9\,arcsec, we
obtained an additional spectrum with a wide slit of 5.0\,arcsec to
minimise slit losses and achieve an absolute flux calibration over the
full wavelength range. Since the seeing at that time was $\sim$0.9\,arcsec,
the resolution of this spectrum is similar to the rest of this dataset.

\subsection{Spectral classification}
\label{SSsclass}

The XSHOOTER spectra (see Fig.~\ref{Fxshoothigh}) 
revealed weak metal lines indicating that WISE~J0725$-$2351
is a metal poor star of spectral type F.
For further reference we use the  
list of Gaia FGK benchmark stars for metallicity
(Jofr{\'e} et al.~\cite{jofre14}, their Table 1), which
includes only three stars with [Fe/H]$<$$-$2.

We retrieved XSHOOTER spectra of two of them from the ESO archive.
The spectrum of the bluest star, HD~84937 (listed in 
Jofr{\'e} et al.~\cite{jofre14} with 
$[Fe/H]$$=$$-$2.08, $T_{eff}$$=$6275\,K, $\log{g}$$=$4.11),
is very similar to that of WISE~J0725$-$2351.
The second best match from their table is 
HD~140283 ($[Fe/H]$$=$$-$2.41, $T_{eff}$$=$5720\,K, $\log{g}$$=$3.67).
These two comparison stars
are also included as the second- and third-brightest objects in the
SIMBAD sample
of sdF subdwarfs with trigonometric parallaxes shown in
Figs.~\ref{Fjksv} and \ref{Fjpmv}. SIMBAD gives very large numbers
of references ($>$500) for both objects and lists a spectral type of
sdF5 for HD~84937, but sdF3 for the cooler HD~140283. Although both objects
have been used as well-investigated standards for a long time, the
catalogue of Skiff~(\cite{skiff14}) lists a large variety of spectral types
between early- and late-F (or -sdF) types (in the case 
of HD~140283 even G-types and luminosity class IV)
from publications between 1966 and 1999 (in some earlier works they were
also classified as A-type (sub)dwarfs).
According to its already mentioned relatively small $\log{g}$ and
the recent interferometric radius measurement of 2.21$\pm$0.08 solar radii
(Creevey et al.~\cite{creevey14}), HD~140283 is a
metal-poor
subgiant rather than a subdwarf.
Even HD~84937, for which a spectral type of sdF5 or F5VI is given
in all recent catalogues listed in VizieR,
is described by VandenBerg et al.~(\cite{vandenberg14})
as a metal-poor turnoff star just beginning its subgiant branch evolution.
All these discrepancies indicate
that there are
problems with a spectroscopic classification scheme at the warm end of
the cool subdwarf sequence.
Therefore, we assign only a preliminary spectral type of sdF5: to
WISE~J0725$-$2351.

\subsection{Radial velocity curve}
\label{SSrv}

To search for radial velocity variations another set of 
medium-resolution spectra ($R$$\simeq$2200, $\lambda$$=$4450-5110\AA{})
were obtained at 13 epochs
with the EFOSC2 spectrograph in February 2014.
Radial velocities ($v_{rad}$) were measured by fitting a set of 
mathematical functions (polynomial, Lorentzian, and Gaussian) 
to the Balmer lines of the EFOSC2 
and XSHOOTER 
spectra using the FITSB2 routine (Napiwotzki et al. 
\cite{napiwotzki04}). No significant variations of $v_{rad}$
were measured within 
the EFOSC2 dataset. The radial velocity is constant at 230$\pm$9\,km/s
(see Fig.~\ref{Frv}), where the average $1\sigma$ error of the single 
measurements is adopted as uncertainty. 
The XSHOOTER spectra in the UVB arm show no variability 
in $v_{rad}$
within $\sim$45\,min. The radial velocity of 238.9$\pm$1.8\,km/s
is perfectly consistent with that derived from the 
EFOSC2 dataset. Since no variations of $v_{rad}$ have been measured 
on timescales of hours, days, and months, we can exclude a close stellar 
companion if its orbit is not aligned in the plane of the sky.

   \begin{figure*}
   \centering
   \includegraphics[width=12.8cm]{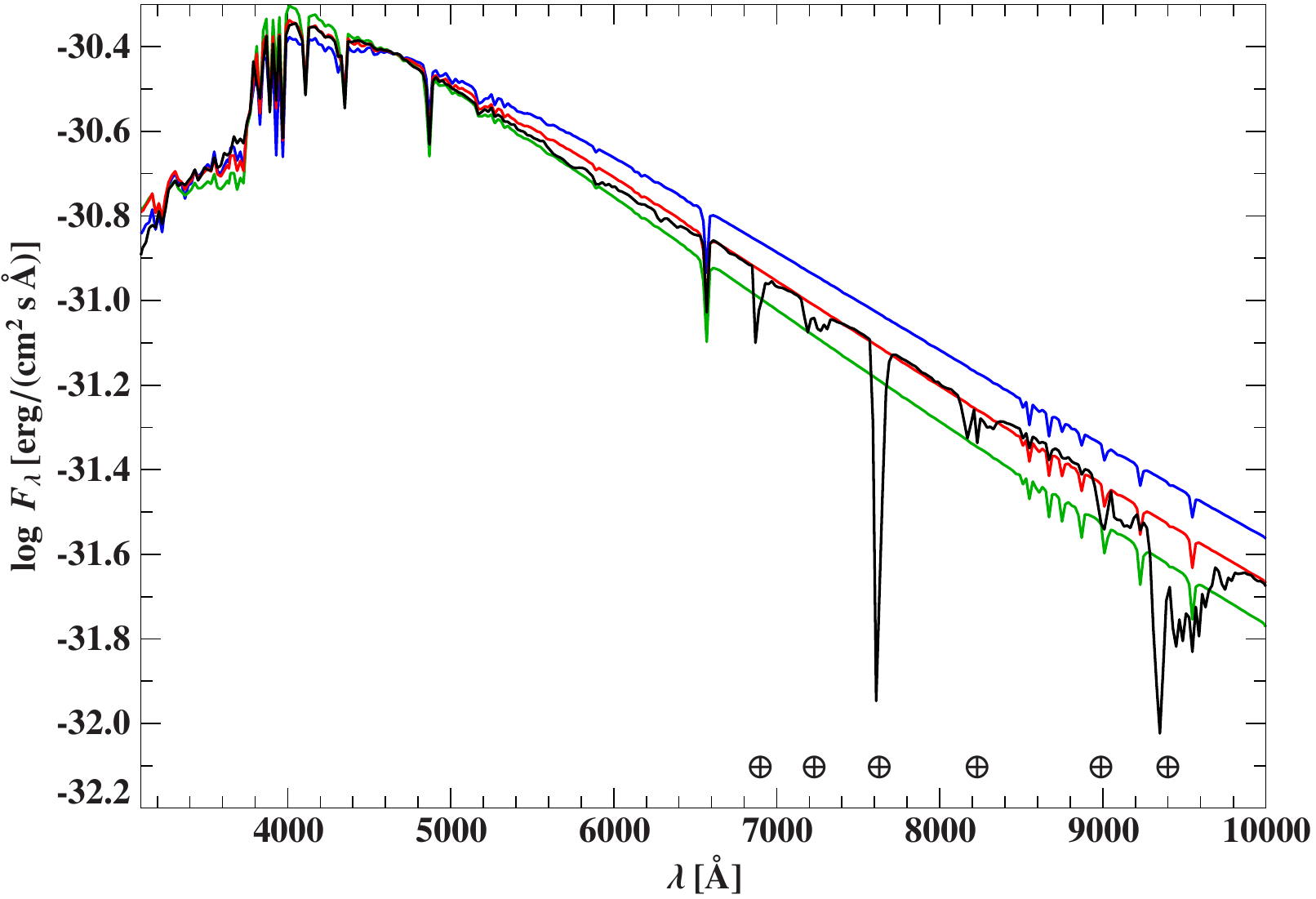}
   \includegraphics[width=12.8cm]{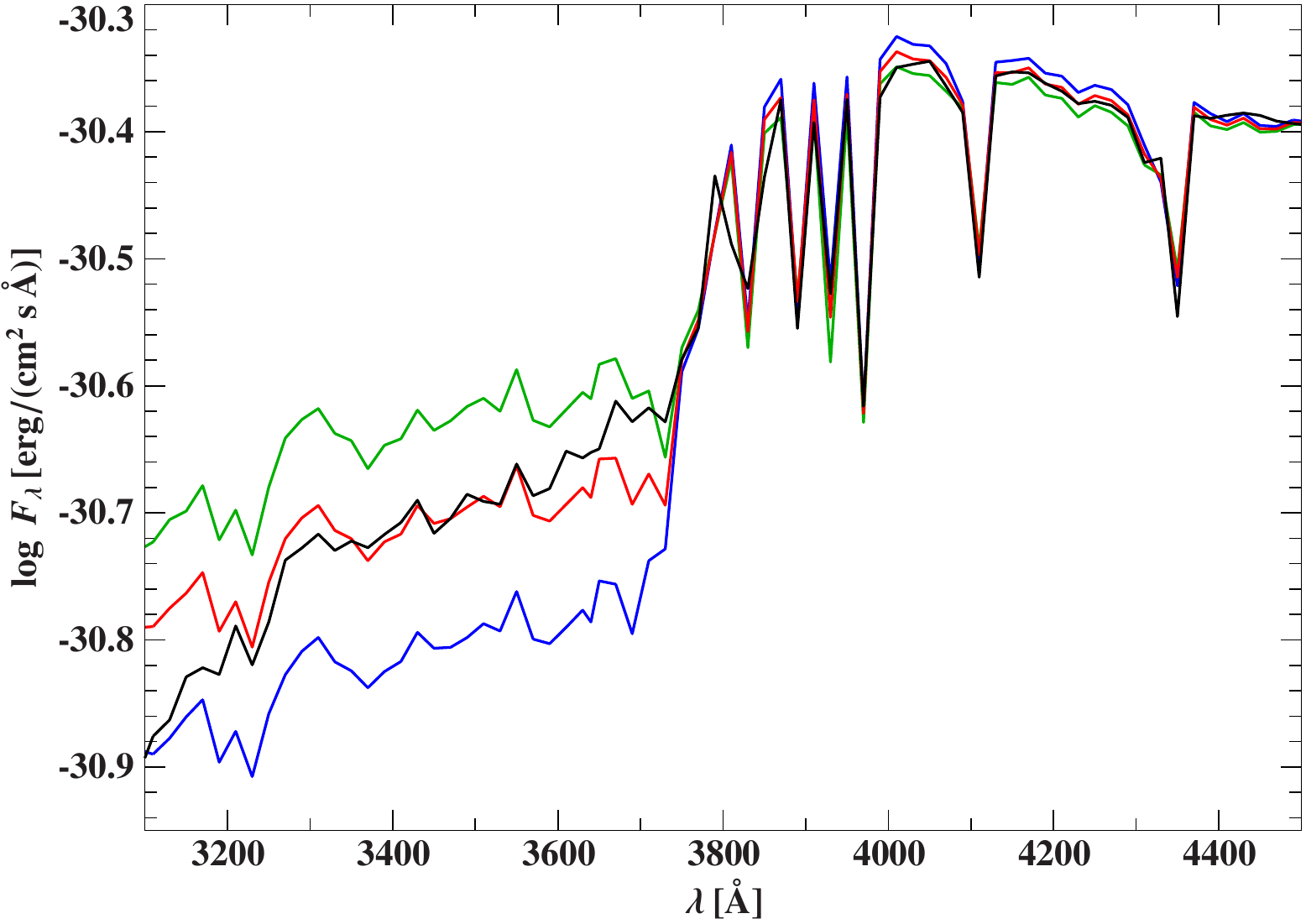}
      \caption{XSHOOTER spectrum of WISE~J0725$-$2351 (black; 
              telluric absorption bands are marked by $\oplus$)
              compared to a synthetic spectrum with [Fe/H]$=$$-$2.0,
              $T_{\rm eff}$$=$6250\,K and $\log{g}$$=$4.0 (red).
              \textbf{Top:} 
              for the full wavelength interval additional synthetic
              spectra with changed temperature $T_{\rm eff}$$=$6500\,K (green) 
              and 6000\,K (blue) are shown.
              \textbf{Bottom:} 
              for the region of the Balmer jump, the additional 
              synthetic spectra show the effect of changing gravity to
              $\log{g}$$=$4.5 (green) and $\log{g}$$=$3.5 (blue).
              }
         \label{Fxshootlow}
   \end{figure*}

   \begin{figure*}
   \centering
   \includegraphics[width=16.3cm]{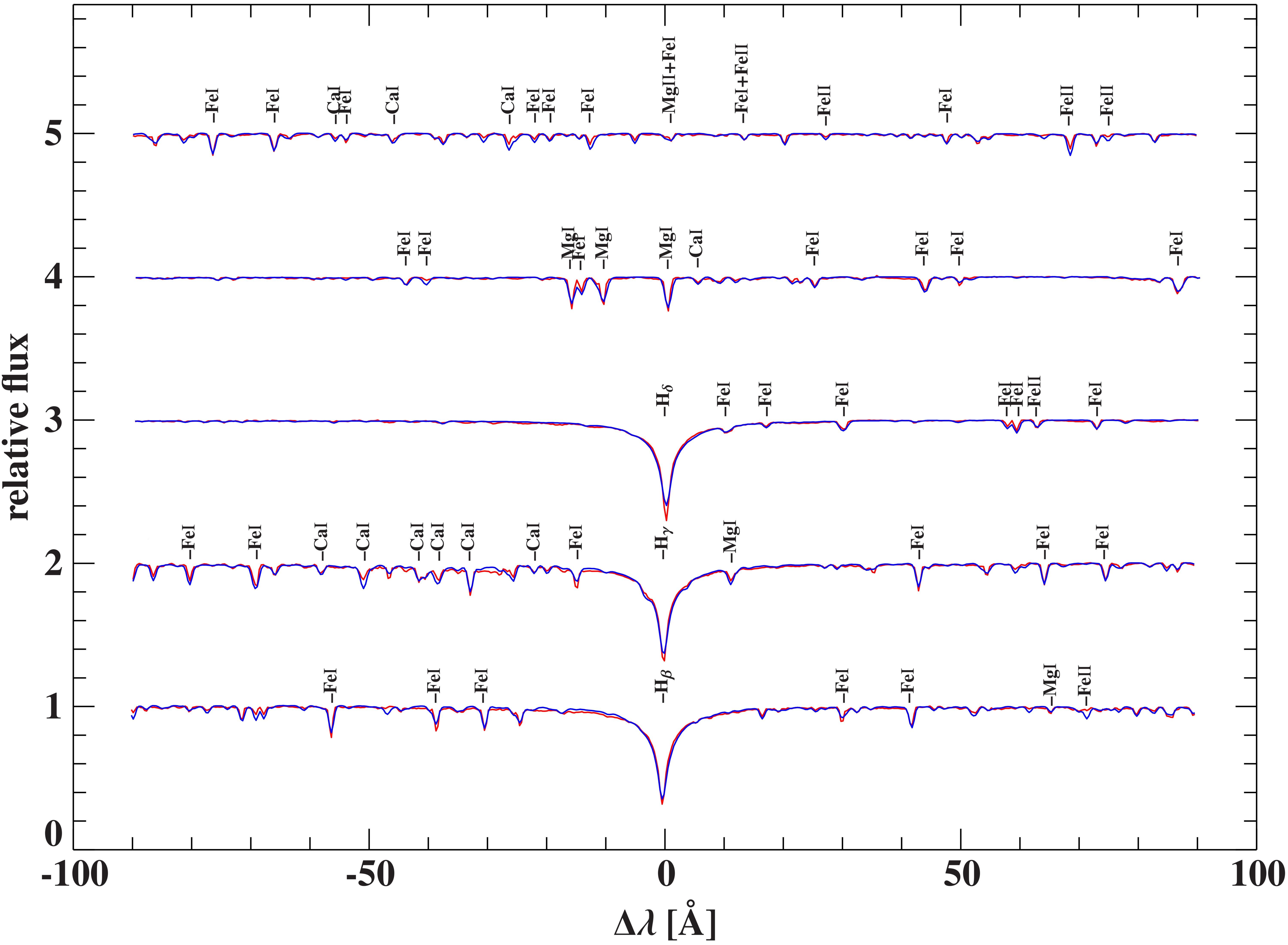}
      \caption{XSHOOTER spectrum of WISE~J0725$-$2351 (red)
              at higher resolution for the regions of the
              Balmer lines H$_{\beta}$, H$_{\gamma}$, and H$_{\delta}$
              and other important spectral lines.
              The best-fit model spectrum with
              $T_{\rm eff}$$=$6250\,K, $\log{g}$$=$4.0,
              [Fe/H]$=$$-$2.0, and [$\alpha$/Fe]$=$$+$0.4 is shown in blue.
              }
         \label{Fxshoothigh}
   \end{figure*}


\section{Quantitative spectral analysis}
\label{Sqspec}

We carried out a quantitative spectral analysis of the XSHOOTER
spectra using grids of Kurucz LTE model atmospheres 
(Castelli \& Kurucz \cite{castelli04}, Munari et al.~\cite{munari05}).

In the first step we compared the flux-calibrated XSHOOTER spectrum 
to synthetic spectra for an adopted metallicity 
of [Fe/H]$=$$-$2.0, an alpha enrichment of [$\alpha$/Fe]$=$$+$0.4 
with a mixing-length parameter of 1.25, and different effective 
temperatures and gravities (see Fig. \ref{Fxshootlow}). The best match is 
found for $T_{\rm eff}$$=$6250\,K and a surface gravity of 
$\log{g}$$=$4.0. As can be seen, the Paschen continuum is a good 
temperature indicator, whereas the gravity can be derived from the 
Balmer jump. We adopted uncertainties of $\delta\,T_{\rm eff}$$=$100\,K
and $\delta\log{g}$$=$0.2\,dex. 
Thereafter, we kept the gravity fixed and proceeded to the analysis of the 
high resolution XSHOOTER spectrum. We used the program FITSB2 
(Napiwotzki et al.~\cite{napiwotzki04})
to fit the Balmer lines H$_{\beta}$, H$_{\gamma}$, and H$_{\delta}$ 
as well as several wavelength ranges covering important spectral lines 
(see VandenBerg et al.~\cite{vandenberg14}). Effective temperature and 
metallicity (with given [$\alpha$/Fe]$=$$+$0.4) were 
derived 
to $T{_{\rm eff}}$$=$6250$\pm$100\,K 
and [Fe/H])$=$$-$2.0$\pm$0.2. 
A comparison of the best-match model spectrum to the XSHOOTER 
observation
is shown in Fig.~\ref{Fxshoothigh}. 
The effective temperature derived is consistent with that derived 
from the flux-calibrated spectrum. 
The errors of our final physical parameters (Table~\ref{Tpar}, 
top rows) were estimated by visual inspection of all fits of different
model spectra to the observed XSHOOTER spectrum.

To verify our procedure, we also studied the halo star 
HD~84937 in the same way. We obtained XSHOOTER spectra from the ESO 
archive. A detailed quantitive spectral analysis has recently been carried 
out by VandenBerg et al.~(\cite{vandenberg14}), which resulted 
in $T{_{\rm eff}}$$=$6408\,K, $\log{g}$$=$4.05, [Fe/H]$=$$-$2.08 
and [$\alpha$/Fe]$=$$+$0.38
(in good aggreement with the parameters in
Jofr{\'e} et al.~\cite{jofre14}.)
Applying the same procedure as for WISE~J0725$-$2351 to the
XSHOOTER spectra of HD~84937, we confirm the results 
of VandenBerg et al.~(\cite{vandenberg14}) to within error limits. 
In particular, the surface gravity derived here from the Balmer jump 
is consistent with that determined by VandenBerg et al.~(\cite{vandenberg14}) 
from mass, $T{_{\rm eff}}$ and bolometric corrections.
In conclusion, WISE~J0725$-$2351 is 
probably a halo turnoff star similar in composition, mass, and age to HD~84937.

%
\begin{table}
\caption{Derived parameters of WISE~J0725$-$2351} 
\label{Tpar}      
\centering                          
\begin{tabular}{lc}        
\hline\hline                 
Parameter  &   WISE~J0725$-$2351 \\ 
\hline                        
Spectral type        & sdF5.0:  \\
$T_{eff}$ [K]        & 6250$\pm$100     \\
$\log{g}$            & 4.0$\pm$0.2      \\
$[Fe/H]$                & $-$2.0$\pm$0.2     \\
\hline
$d_{phot}$ [pc]      & 430$^{+120}_{-90}$ \\
$v_{tan}$ [km/s]     & 544$^{+151}_{-114}$    \\
$v_{rad}$ [km/s]     & $+$238.9$\pm$1.8      \\
\hline
$v_{grf}$ [km/s] & 460$^{+135}_{-102}$ \\ 
$u$ [km/s]           & $-$247$^{+95}_{-60}$ \\
$v$ [km/s]           & $-$165$^{+57}_{-43}$ \\
$w$ [km/s]           & $-$351$^{+85}_{-63}$ \\
\hline                                   
\end{tabular}
\end{table}


\section{Photometric distance and space velocity}
\label{Sdistv}

\begin{figure}
   \centering
   \includegraphics[width=8.8cm]{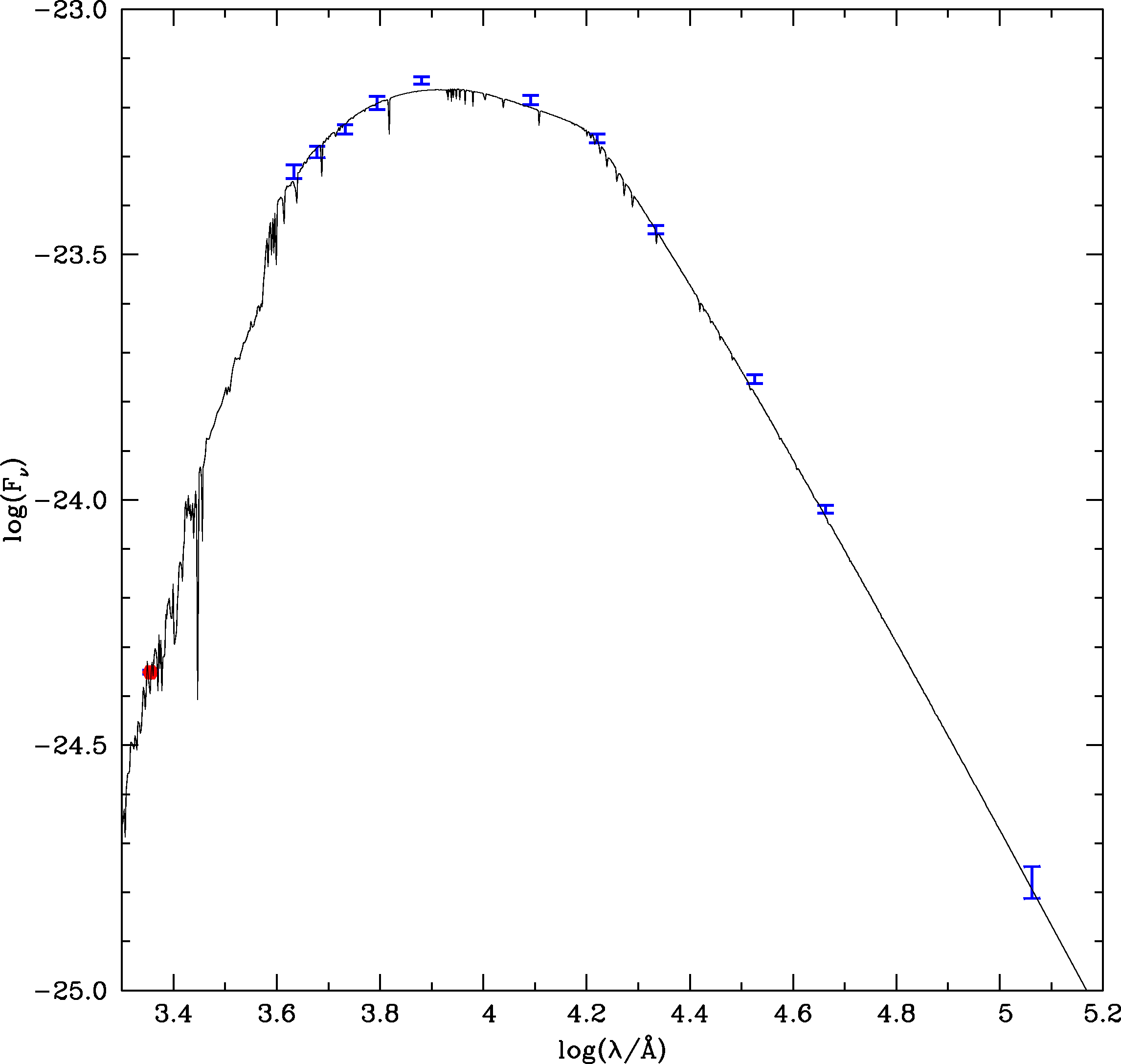}
      \caption{Comparison of the observed flux distribution of
               WISE~J0725$-$2351 derived from the photometry listed in
               Table~\ref{Tphot} and dereddened for $E(B-V)$$=$0.03\,mag
               to a synthetic spectrum with $T{_{\rm eff}}$$=$6250\,K,
               $\log{g}$$=$4.0 and  [Fe/H]$=$$-$2.0. The GALEX NUV flux
               is displayed a red hexagon for clarity. Its uncertainty
               is much lower than the symbol size.
               }
         \label{Fsed}
   \end{figure}

Finally, we inspected the spectral energy distribution (SED) by
making use of the photometric measurements summarised in Table~\ref{Tphot}
thereby covering the SED from the near-ultraviolet (NUV) (GALEX) to the 
far-infrared (FIR) (WISE).
We used the calibrations of
Morrissey et al.~(\cite{morrissey07}),
Pickles \& Depagne~(\cite{pickles10}),
Cohen et al.~(\cite{cohen03}), and
Jarrett et al.~(\cite{jarrett11})
to convert magnitudes to fluxes. We adopted the atmospheric parameters derived
above and included interstellar absorption. The magnitudes were dereddened
using the extinction cofficients of
Yuan et al.~(\cite{yuan13}). 
The synthetic spectra were scaled to match the
$K_s$-band flux.
The best match is derived for a colour access of $E(B-V)$$=$0.03\,mag,
indicating very little extinciton to WISE~J0725$-$2351.
As can be seen from Fig.~\ref{Fsed}, the synthetic spectrum matches
the observed SED perfectly.
The fit also allowed us to determine the distance because the angular diameter
was derived from the scaling factor. In addition, the stellar radius can
be derived from the gravity and the stellar mass.
We adopted 0.8\,M$_\odot$.
Accordingly, the distance was determined as 430$^{+120}_{-90}$ pc, where
the uncertainties stem from the uncertainty of the gravity.

Because of very good agreement in the physical parameters of our
target with those of HD~84937, we also
used the relatively well-known \textit{Hipparcos} 
distance (72.8$\pm$4.1\,pc according to van Leeuwen~\cite{vanleeuwen07}) 
of HD~84937 for an additional
photometric distance estimate of WISE~J0725$-$2351.
Using the $BVJHK_sw2w3$ photometry, and conservatively assuming the absolute 
magnitude uncertainties of about $\pm$0.2\,mag, we estimated a mean photometric
distance of 391$\pm$39\,pc. 
If we prefer the smaller and more accurate parallax of HD~84937
recently measured by VandenBerg et al.~(\cite{vandenberg14}) using the
\textit{Hubble Space Telescope}, the distance
of WISE~J0725$-$2351 
increases to 439$\pm$44\,pc. These distance estimates based
on the assumption that WISE~J0725$-$2351 is just a more distant copy of
HD~84937 and neglecting possible differences in the reddening of the two
objects are in very good agreement with the distance derived from the 
SED. We adopt the distance derived from the SED with its larger errors 
until the gravity 
of WISE~J0725$-$2351 will be determined with higher accuracy.

With this distance, the proper motion of WISE~J0725$-$2351 results in
an extremely large tangential velocity 
(Table.~\ref{Tpar}). 
This is however still consistent with
the majority of the known sdF/sdG subdwarfs having tangential velocities
between 100 and up to about 500\,km/s (Fig.~\ref{Fjpmv}, bottom).
Despite its larger uncertainty, the tangential velocity is clearly larger 
than the radial velocity. The object is 
located at $l$$\sim$238.0$^{\circ}$, $b$$\sim$$-$3.6$^{\circ}$, right in the
Galactic plane.  

We calculated the 
Galactocentric
kinematic properties of WISE~J0725$-$2351 
(bottom rows in Table~\ref{Tpar}) based 
on the input parameters given in Tables~\ref{Tpm} 
and \ref{Tpar} (middle part) following 
the equations of Randall et al.~(\cite{randall14})
by varying the position and velocity components within their 
respective errors by applying a Monte Carlo procedure with a depth of 100000. 
The distance of the sun 
from the Galactic center was assumed to be 8.4 kpc. According to
Sch{\"o}nrich et al.~(\cite{schoenrich10}) its motion relative to the 
local standard of rest (LSR) is $v_{x\odot}$$=$11.1\,km/s, 
$v_{y\odot}$$=$12.24\,km/s, $v_{z\odot}$$=$7.25\,km/s, and the velocity 
of the LSR is $v_\text{LSR}$$=$242\,km/s, as predicted by Model I of 
Irrgang et al.~(\cite{irrgang13}). We derived a Galactic restframe velocity 
of WISE J0725-2351 
$v_{grf}$$=$460$^{+135}_{-102}$\,km/s, 
a Galactic radial velocity component $u$$=$$-$247$^{+95}_{-60}$\,km/s, 
a rotational component $v$$=$$-$165$^{+57}_{-43}$\,km/s, and a component 
$w$$=$$-$351$^{+85}_{-63}$\,km/s perpendicular to the Galactic plane. 
These values imply that the star is a halo star on a retrograde bound orbit 
(about 9\% of the simulations led to unbound orbits)
that is passing close by the Galactic disc. Its location in a $u$-$v$ diagram 
(see e.g. Fig.~2 in Pauli et al.~\cite{pauli06} for WDs, 
or Fig.~10 in Tillich et al.~\cite{tillich11} for sdB stars)
is clearly outside the limits of the thin and thick disc populations.
The $w$ component is very large compared to those of 
M-type halo subdwarfs in L{\'e}pine et al.~(\cite{lepine03}).


\section{Discussion and conclusions}
\label{Sdiscu}

We have discovered a new F-type subdwarf (sdF5:)
or metal-poor turnoff star, which is very similar to the case of HD~84937, 
one of the
best-known representatives of this elusive class of objects. The new
object, WISE~J0725$-$2351, is currently located at about 400\,pc 
from the sun in the Galactic plane,
but crosses it at a high speed typical of an extreme 
Galactic halo object.
The velocity component perpendicular to the plane is very large, 
and the 
negative Galactic rotational velocity component indicates
a retrograde orbit.
We expect WISE~J0725$-$2351 to be roughly the same age as HD~84937,
for which VandenBerg et al.~(\cite{vandenberg14}) determined an age of
about 12 Gyr. 
With its $\log{T{_{\rm eff}}}$$=$3.796 and the (uncertain)
absolute magnitude of $M_V$$=$3.89, WISE~J0725$-$2351 would be placed
in between HD~84937 and HD~19445, but closer in absolute magnitude to
HD~84937, as shown in Fig.~1 of VandenBerg et al.~(\cite{vandenberg14}).
These authors measured 
$[Fe/H]$$=$$-$2.03, $T_{eff}$$=$6136\,K, and $\log{g}$$=$4.43 for
HD~19445, a well-measured
subdwarf about $\sim$1\,mag below the turnoff at the metallicity 
of $[Fe/H]$$\sim$$-$2, which they included in their analysis 
to check the isochrones. According to our gravity
measurement for WISE~J0725$-$2351, it seems more likely to be a 
metal-poor turnoff star than a normal F-type subdwarf.

We classify WISE~J0725$-$2351 as a relatively metal-poor star
at the boundary to the class of extremely metal-poor stars. Its 
colours meet only three out of seven selection
criteria for the best and brightest metal-poor stars as suggested
by Schlaufman \& Casey~(\cite{schlaufman14}). In particular, its 
$J$$-$$H$$=$0.27 and $J$$-$$w2$$=$0.35 are not as red as the required
limits (0.45 and 0.5, respectively), whereas its $w1$$-$$w2$$=$$-$0.03
and $B$$-$$V$$=$0.43 are in the right ranges. 
In that respect, WISE~J0725$-$2351 is again similar to the
benchmark metal-poor subdwarf or turnoff star 
HD~84937. The other two metal-poor
Gaia benchmark stars from Jofr{\'e} et al.~(\cite{jofre14}), the 
already mentioned slightly cooler subgiant HD~140283 and especially the 
much cooler giant HD~122563
($[Fe/H]$$=$$-$2.59, $T_{eff}$$=$4608\,K, $\log{g}$$=$1.61),
show larger $J$$-$$H$ and $J$$-$$w2$ colour indices so that the latter
fulfils most of the Schlaufman \& Casey~(\cite{schlaufman14}) preconditions.
Nevertheless, we think that
WISE~J0725$-$2351 is a good target for higher resolution spectroscopy
and the analysis of elemental abundances.


\begin{acknowledgements}
E.Z. and C.H. acknowledge support by the Deutsche 
Forschungsgemeinschaft
(DFG) through grants HE\,1356/45-2 and HE\,1356/62-1, respectively.
We thank U. Munari for providing us with his synthetic spectra of high 
spectral resolution.  
This research has made use of the National Aeronautics and Space Administration
(NASA)/Infrared Processing and Analysis Center (IPAC) Infrared Science Archive, which is operated by the Jet Propulsion
Laboratory (JPL), California Institute of Technology (Caltech), 
under contract with the NASA, of data products from WISE,
which is a joint project of the University of California,
Los Angeles, and the JPL/Caltech, funded by the NASA, and from 2MASS.
We have extensively used SIMBAD and VizieR at the CDS/Strasbourg.
We thank the anonymous referee for a prompt report.
\end{acknowledgements}


\end{document}